\def\ga{\mathrel{\raise.3ex\hbox{$>$\kern-.75em\lower1ex\hbox{$\sim$}}}}
\def\la{\mathrel{\raise.3ex\hbox{$<$\kern-.75em\lower1ex\hbox{$\sim$}}}}
\def\ep{\epsilon}
\def\order#1{{\cal O}\left(#1\right)}
\def\ba{\begin{eqnarray}}
\def\ea{\end{eqnarray}}
\def\eq#1{(\ref{#1})}
\newcommand{\Wsla}{W\hspace{-0.7em}/}

\newcommand{\fr}[2]{\frac{#1}{#2}}

\documentstyle[12pt,epsf,psfig,epsfig]{article}

\begin{document}


\hspace{\fill} Alberta Thy 11-04 \rightline{UVIC--TH--04/07}

\begin{center}

\large {\bf Electric Dipole Moments of Leptons \\[2mm]
in the Presence of Majorana Neutrinos} \normalsize

\vspace{1cm}

John Paul Archambault$^1$,  Andrzej Czarnecki$^1$
 and
Maxim Pospelov$^2$

\vspace{1cm}

$^1${\it Department of Physics, University of Alberta, \\Edmonton,
AB, T6G 2J1 Canada}\\
$^2${\it Department of Physics and Astronomy, University of Victoria, \\
     Victoria, BC, V8P 1A1 Canada}

\end{center}


\begin{abstract}

We calculate the two-loop diagrams that give a non-zero
contribution to the electric dipole moment $d_l$ of a charged
lepton $l$ due to possible Majorana masses of neutrinos. Using the
example with one generation of the Standard Model leptons and two
heavy right-handed neutrinos, we demonstrate that the
non-vanishing result for $d_l$ first appears in order $O(m_l
m_\nu^2 G_F^2)$, where $m_\nu$ is the mass of the light neutrino
and the see-saw type relation is imposed. This effect is beyond
the reach of presently planned experiments.
\end{abstract}

\vspace{4cm}

\hfill\eject

\section{Introduction}

Recent discovery of CP violation in the neutral $B$-meson system
\cite{Babar,Belle} is in perfect accord with the CP violation
observed in $K$ mesons \cite{1964}. Both results are explained
within the minimal model of CP violation, known as the
Kobayashi-Maskawa model \cite{KM}. It links CP-violation in $K$
and $B$ mesons to a single CP-violating invariant of the
Kobayashi-Maskawa matrix in the quark sector,
Im$(V_{tb}V^*_{td}V_{cd}V^*_{cb}) \simeq 3 \times 10^{-5}$. This
combination, as well as $\theta_{\rm QCD}$, are the only sources
of CP violation in the Standard Model.

An independent piece of experimental information about CP
violation comes from the searches of the electric dipole moments
(EDMs) of neutrons \cite{n} and heavy atoms \cite{Tl,Hg}. It is
well known that the Kobayashi-Maskawa model predicts extremely small
EDMs. Indeed, the necessity of four electroweak vertices requires
from any diagram capable of inducing an EDM of a quark to have at
least two loops. Moreover, it turns out that all EDMs or color
EDMs of quarks vanish exactly at the two-loop level \cite{Shab},
and only three-loop diagrams survive \cite{Kh,CK}, producing a
tiny number of order $10^{-34}e~$cm.

The only relevant operator that is not zero at the two-loop order
is the Weinberg operator \cite{W-P}, but its numerical value turns
out to be also extremely small. Possible enhancement comes from
the large distance effects, that could lead to a KM-generated EDM
of the neutron of order $10^{-32}e~$cm \cite{LD}, which is still
six to seven orders of magnitude smaller than the current
experimental limit. The KM phase in the quark sector induces the
EDM of a lepton via a diagram with a closed quark loop, but a
non-vanishing result appears first at a four-loop level \cite{KP}
and therefore is even more suppressed.

The suppressed values of EDMs from the KM model together with
enormous accuracy of EDM experiments produce stringent constraints
on possible  flavour-diagonal sources of CP violation, notably on
CP-odd combinations of the soft-breaking parameters in the
supersymmetric extensions of the Standard Model (SM).

On the other hand, the KM phase cannot be the only source of CP
violation in nature. Dynamical generation of the baryon asymmetry
of the Universe (BAU) requires presence of an additional source(s)
of CP violation in nature. One of the most appealing scenarios for
BAU is leptogenesis \cite{Leptog} where a non-zero lepton number
is generated from the out-of-equilibrium decay of heavy
right-handed Majorana neutrinos with CP violation coming from the
Yukawa sector of the theory. Subsequent sphaleron transitions
\cite{KRS} transform half of the initial lepton asymmetry into
BAU.

The attractiveness of leptogenesis is in its simplicity, relative
freedom from the low-energy constraints, and in recent
experimental results that indicate large mixing angles in the
lepton sector \cite{nuosc}.

The complex phases in the lepton mixing suggested by leptogenesis
at some level will induce $d_e$, the electron EDM, as well as EDMs
of other charged leptons. Various calculations of $d_e$ and
$d_\mu$ were performed in the supersymmetric case under certain
assumption about the soft-breaking parameters \cite{susy}, while a
non-supersymmetric case remains poorly explored.

We note that manifestations of Majorana phases in CP-violating
phenomena has been explored in \cite{deGouvea:2002gf}.  The EDM is
an additional observable arising due to those phases.  Compared
with the KM model, where three generations and at least four loops
are necessary to generate a lepton EDM, the addition of Majorana
phases enhances the effect.  However, as we will see, it remains
very small.

\begin{figure}[htb]
\hspace*{25mm}
\begin{tabular}{c@{\hspace*{10mm}}c}
\epsfxsize=45mm \epsfbox{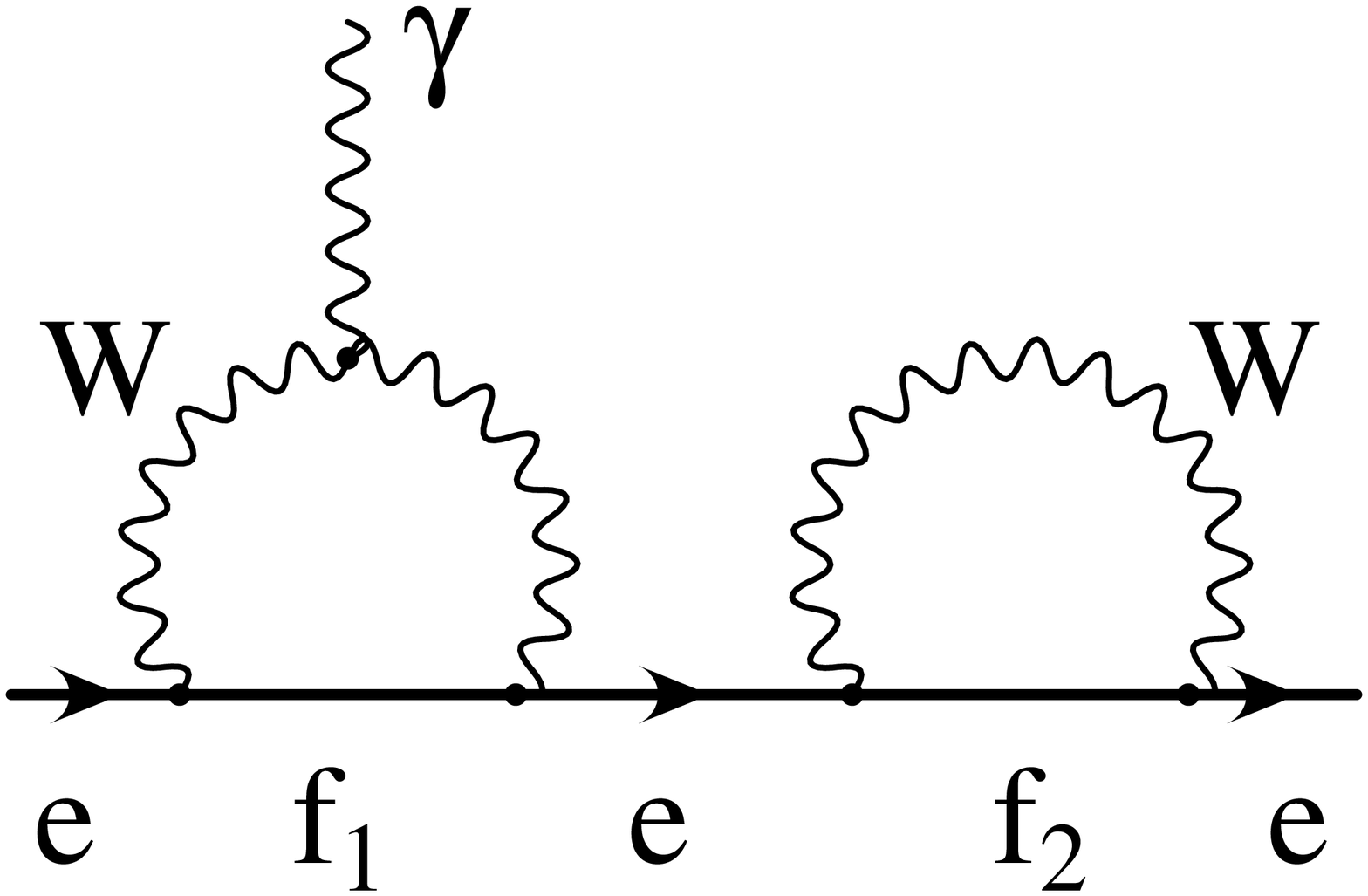} & \epsfxsize=45mm
\epsfbox{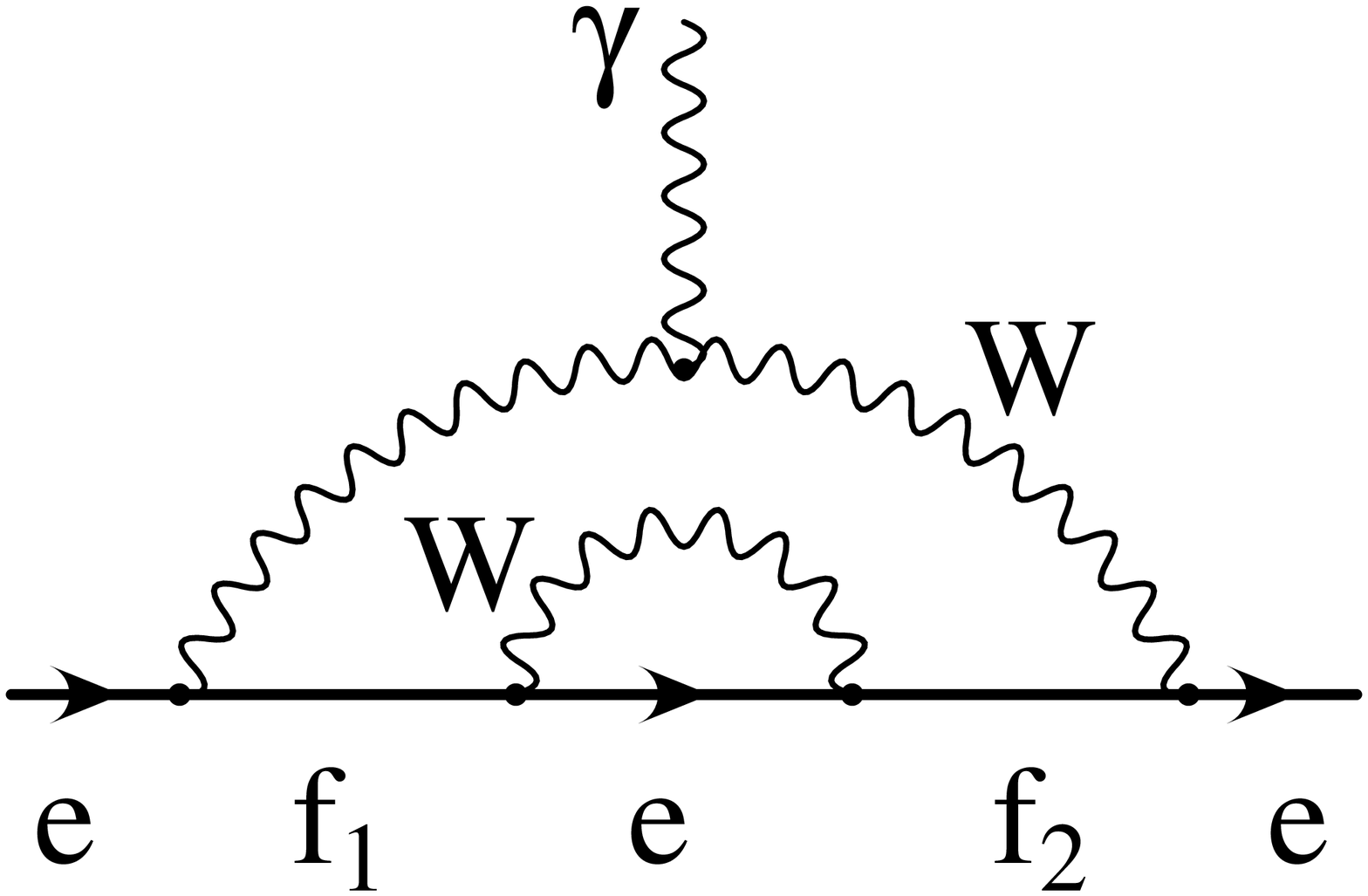}
\\
(a) & (b)
\end{tabular}
\caption{Examples of two-loop diagrams that give zero electric
dipole moment of the electron. The direction of the arrows remains the
same on all fermion propagators and the external photon can be
attached to any charged line.} \label{figDirac}
\end{figure}

In this paper, we present a systematic analysis of the two-loop
diagrams that lead to the EDMs of charged leptons through
interactions with two additional heavy neutrinos, in a
one-generation model suggested by a minimal leptogenesis scenario.
These diagrams can be divided into two major classes. The first
class where the lepton number is conserved along the fermion line
vanishes identically, as the same arguments that lead to $d_q=0$
at two loops \cite{Shab} apply (see Fig. \ref{figDirac}).

\begin{figure}[htb]
\hspace*{0mm}
\begin{tabular}{c@{\hspace*{10mm}}c@{\hspace*{10mm}}c}
\epsfxsize=45mm \epsfbox{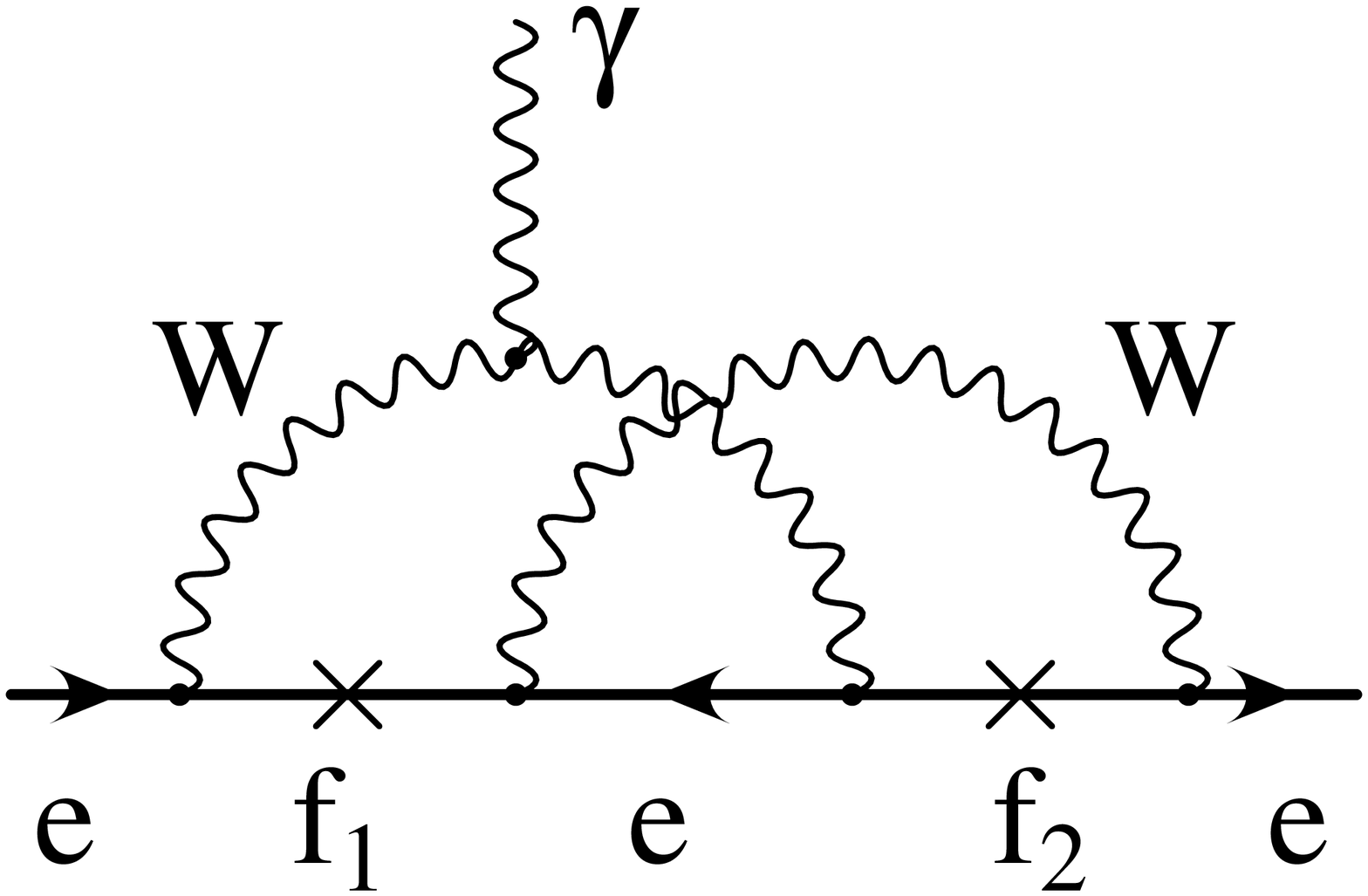} & \epsfxsize=45mm
\epsfbox{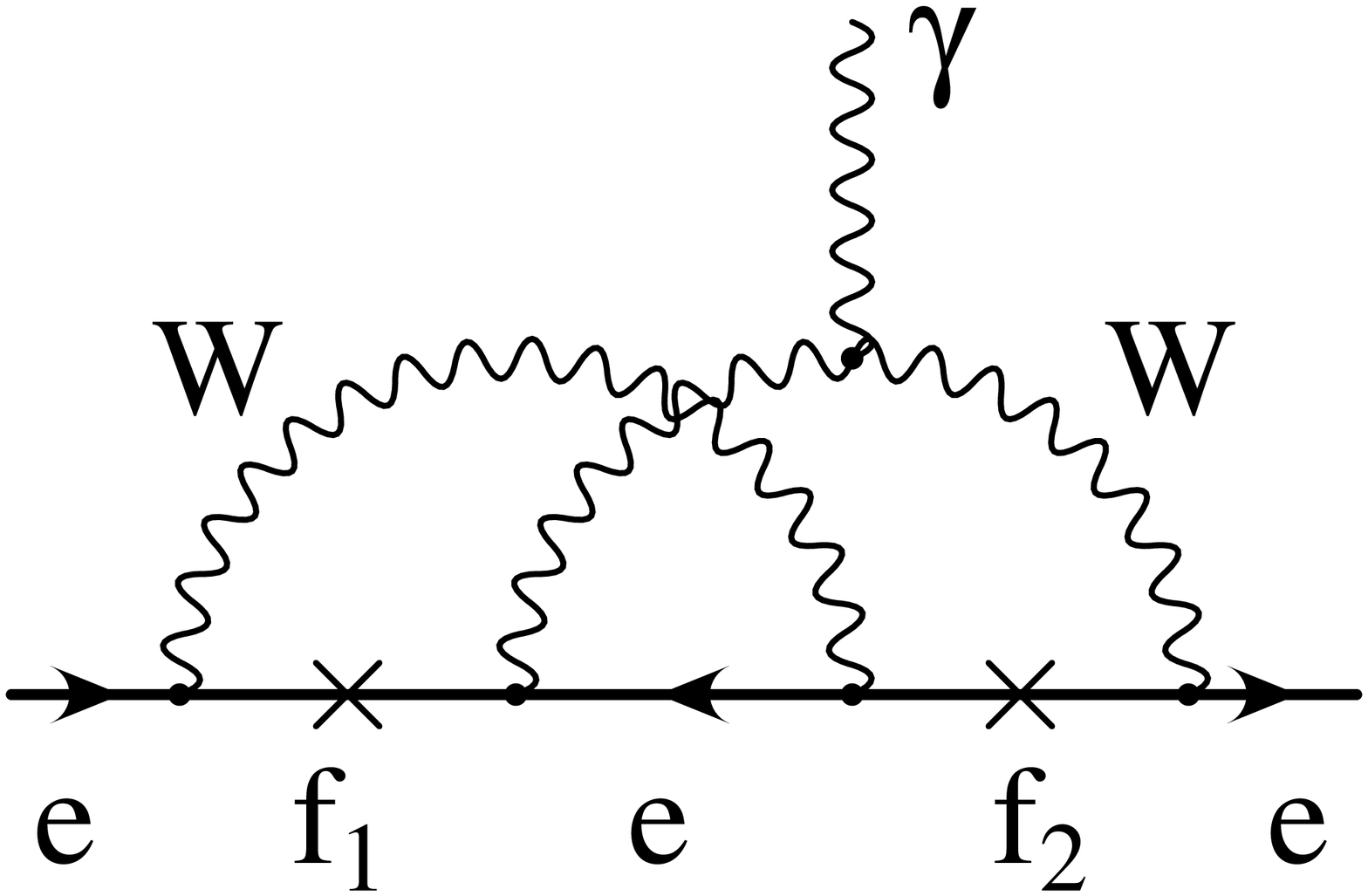} & \epsfxsize=45mm \epsfbox{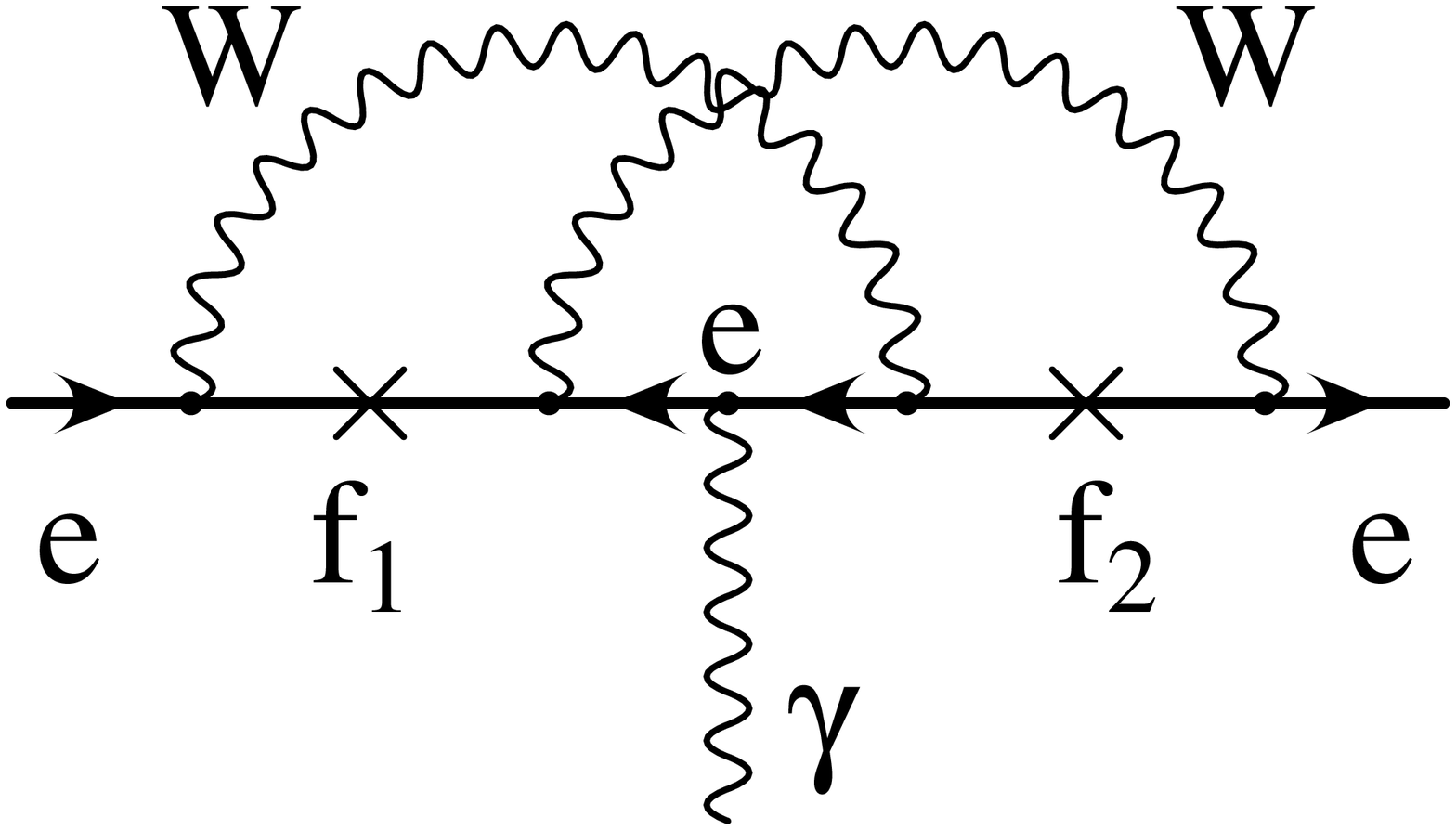}
\\
(1) & (2) & (3)
\end{tabular}
\caption{Contributions to the electron EDM in a model with
Majorana masses of neutrinos.  $f_{1,2}$ denote all possible
neutrinos (see text).  Crosses denote insertions of lepton-number
violating mass parameters. Note that the direction of the internal
electron line is opposite to the external ones.} \label{edm}
\end{figure}

The second class of diagrams  (Fig. \ref{edm}) has no analogues in
the quark sector: due to the existence of the Majorana component
of the neutrino mass matrix, a combination of $\Delta L = 1$ and
$\Delta L = -1$ transitions is possible. The existence of this
additional class of diagrams for leptons was pointed out in Ref.
\cite{Ng^2}. It leads to a non-zero electron EDM ($d_e$) even at the
two-loop level, but up to now there has not been any serious attempt
to calculate the size of $d_e$ induced by this class of diagrams. (An
estimate of $d_e$ due to CP violation in the sector of heavy Majorana
neutrino was presented in Ref. \cite{Pilaftsis}.  However, this result
is not satisfactory, as it suggests that $d_e$ does not vanish in the
limit of infinitely heavy right-handed neutrinos.)

The purpose of this work is to give a detailed calculation of
$d_e$ due to the CP violation in the lepton sector in the presence
of Majorana masses for neutrinos and obtain a prediction for the
electron EDM in the see-saw model of the neutrino mass sector.
A similar mechanism can of course also lead to the EDM of the muon.

We find that two-loop diagrams give non-vanishing results for the
$d_e$.  However, EDMs are very much suppressed by the smallness of
the neutrino mass as they first appear in $O(m_em_\nu^2G_F^2)$
order, if the see-saw type relation between $m_\nu$, Dirac mass
and heavy Majorana mass is imposed. Thus, for any
phenomenologically motivated choice of $m_\nu$ the result for
$d_e$ (and $d_\mu$)  turns out to be much smaller than current or
projected experimental sensitivity to EDMs. We also note that the
fine-tuning in the neutrino mass sector may lead to up to ten
orders of magnitude enhancement in $d_e$, which is still not
enough to bring it within the experimental reach in the near
future.

\section{Description of the model}

We take one standard model generation: $\left(
\begin{array}{c}\nu_L\\ e_L \end{array} \right)$, $e_R$, and two
singlet heavy neutrinos $N_{1,2}$. The latter do not participate
in electroweak interactions; in particular, the charged current
sector is described by the Lagrangian
\ba
{\cal L}_{cc} = {g\over
\sqrt{2}}\left( \bar \nu_L \Wsla^+ e_L \, + \,
\mbox{(H.c.)}\right) \label{cc}
\ea

The mass sector Lagrangian for fermions is
\ba
-{\cal L}_M &=& m_e
\left( \bar e_L e_R  +  \bar e_R e_L \right)
\nonumber \\
&& +{M_1\over 2}\left( \bar N^c_1 N_1 + \bar N_1 N_1^c \right)
+{M_2\over 2}\left( \bar N^c_2 N_2 + \bar N_2 N_2^c \right)
\nonumber \\
&&
 +m_1\left( e^{i\phi_1} \bar N_1 \nu_L + e^{-i\phi_1} \bar \nu_L N_1 \right)
\nonumber \\
&& +m_2 \left( e^{i\phi_2} \bar N_2 \nu_L +e^{-i\phi_2} \bar \nu_L
N_2 \right).
\ea
Here $\psi^c \equiv \gamma_0 C\psi^*$; $M_{1,2}$
and $m_{1,2}$ are defined in terms of real positive Yukawa
couplings $y_{1,2}$ and the electroweak vacuum expectation value
$v$,
\ba m_{1,2} \equiv {y_{1,2} v\over \sqrt{2}}.
\ea
We use the
freedom of phase choice for $\nu_L$ and $e_{R,L}$ to redefine
\ba
 \nu_L \to e^{-i\phi_2}\nu_L.
\ea
We see that there is only one physical CP violating phase
$\eta\equiv \phi_1 - \phi_2$.

Before we explore the physical manifestation of $\eta$, we
determine the mass eigenstates of neutrinos.  We will use them to
compute the EDM of the electron. We use the identity
\ba \bar
\nu_L N &=& {1\over 2}\left(\bar \nu_L N + \bar N^c \nu_L^c\right)
\nonumber \\
\bar N \nu_L  &=& {1\over 2}\left(\bar N \nu_L + \bar  \nu_L^c
N^c\right)
\ea
to rewrite the neutrino mass matrix as
\ba
 -{\cal
L}_{M,\nu} &=& {1\over 2}(\bar \nu_L, \bar N_1^c, \bar N_2^c)
{\cal M} \left(
\begin{array}{c}
\nu_L^c \\
N_1\\
N_2
\end{array}
\right) + \mbox{(H.c.)}
\nonumber \\
{\cal M} &\equiv & \left(
\begin{array}{ccc}
0  &  m_1 e^{i\eta} & m_2 \\
m_1 e^{i\eta} & M_1 & 0 \\
m_2 & 0 & M_2
\end{array}
\right)
\label{massmat}
\ea
We use $M_{1,2} \gg m_{1,2}$ to approximately
diagonalize the neutrino mass matrix. We define neutrino mass
eigenstates: light $\nu$ and heavy $n_{1,2}$, \ba \left(
\begin{array}{c}
\nu_L^c \\
N_1\\
N_2
\end{array}
\right) = V \left(
\begin{array}{c}
\nu^c \\
n_1\\
n_2
\end{array}
\right) \label{mix} \ea where $V$ is the matrix diagonalizing
${\cal M}$.  Approximately, \ba V&=& \left(
\begin{array}{ccc}
1 &  {m_1 e^{i\eta} \over M_1} & {m_2\over M_2} \\
-{m_1 e^{i\eta}\over M_1} & 1 & 0 \\
-{m_2\over M_2} & 0 & 1
\end{array}
\right)
\nonumber \\
V^T {\cal M} V&=& \left(
\begin{array}{ccc}
 - {m_1^2 e^{2i\eta} \over M_1} - {m_2^2 \over M_2} & o_2 & o_2\\
o_2 & M_1+o_1 & o_1\\
o_2 & o_1 & M_2 + o_1
\end{array}
\right) \ea
where we denote small corrections by $o_n \equiv
\order{m\left({m\over M}\right)^n}$. Two things are worth noting:
\begin{itemize}
\item Heavy neutrinos $n_{1,2}$ participate in charged current
interactions described by eq.~\eq{cc}, through their presence in
$\nu_L$, eq.~\eq{mix}:
\ba \nu_L^c\simeq \nu^c +{m_1
e^{i\eta}\over M_1}n_1 + {m_2\over M_2}n_2.
\ea
\item The CP
violating phase enters through the complex mass of $\nu$ and
through the charged current interaction of $n_1$.
\end{itemize}
The Majorana mass of the light (active) neutrino is given by the
following relation,
\ba m_{\nu} = \left| {m_1^2
e^{2i\eta} \over M_1} + {m_2^2 \over M_2} \right|.
\label{mnu}
\ea
Since experiments with light neutrinos give information about
$m_\nu$, it is convenient to keep this parameter fixed, while
allowing $m_{1,2}$, $M_{1,2}$ and $\eta$ to vary. In doing this,
we would have to distinguish two possibilities.
\begin{itemize}
\item See-saw relation, when both contributions in (\ref{mnu}) are
on the order or smaller than $m_\nu$, \item Cancellation of two
terms in $m_\nu$ which we would term as a "fine-tuned case",
although it could be a result of some symmetry that suppresses the
determinant of the mass matrix (\ref{massmat}). For this
cancellation to happen, one has to have $m_1^2/M_1 \simeq
m_2^2/M_2$ and phase $\eta$ close to $\pi/2$.
\end{itemize}
We would like to note that theoretically one could have a
fine-tuned case even for $m_{1,2} \sim M_{1,2}$. This, however,
would also result in large admixtures of heavy $n_1$ and $n_2$
neutrinos in the original left-handed neutrino. In a more
realistic model with three active neutrino species, such
admixtures could lead to a non-universality of charged currents,
modification of the invisible $Z$ decay width, etc.  With the
accuracy of electroweak data, we impose the bound on the size of
the mixing angle: \ba \fr{m_1}{M_1},~\fr{m_2}{M_2}\la
\order{1\over 10}. \label{univ} \ea

\section{Computation of the electron EDM at two loops}

To define the electron EDM, consider the general matrix element of
an electromagnetic current between spin 1/2 fermions,
\ba &&\bar
u_f(p) \left\{ F_1(t) \gamma_\mu - {i\over 2m} F_2(t)
\sigma_{\mu\nu} q^\nu
                        +{1\over m} F_3(t)q_\mu
                        \right.
 \nonumber\\
&&\qquad \left. +\gamma_5 \left[
 G_1(t) \gamma_\mu - {i\over 2m} G_2(t) \sigma_{\mu\nu} q^\nu +{1\over m}
G_3(t)q_\mu
 \right]\right\}  u_i(p+q),
\ea
with $t\equiv q^2$.  The EDM $d_e$ of an electron of charge
$e$ is given by
\ba d_e= -{ie\over 2m_e} G_2(0). \ea

The most notable property of all diagrams in Figs. \ref{figDirac}
and \ref{edm}, potentially contributing to the CP-odd amplitude,
is the effective antisymmetrization over the neutrino propagation,
$f_1$ and $f_2$. This is because the CP-odd part of a diagram with
the selection of eigenstates $f_1$ and $f_2$ is always {\em
opposite} to the CP-odd part of the diagram with interchanged
flavours. As a result of this antisymmetrization, in the expansion
over small $q$  the first class of diagrams, Fig. \ref{figDirac}a
and \ref{figDirac}b, turns out to be proportional to the cube of
the photon momentum $O(q^3)$, while $G_2(0)$ vanishes identically.
A more detailed explanation of the cancellation of EDMs can be
found in Ref. \cite{Shab,KP,W-P}.

Within the model we are considering, the non-vanishing diagrams
contributing to $d_e$ are shown in Fig.~\ref{edm}. Their
evaluation is simplified if we assume the most natural mass
hierarchy, ${m_{1,2}^2\over M_{1,2}}\ll m_e \ll M_W \ll M_{1,2}$.
We use the notation $M={M_1+M_2\over 2}$ and $\Delta M={M_2-M_1}$
and assume $|\Delta M| \ll M$ for calculational convenience.

First, we treat the case when $f_{1,2}$ are both heavy (some
details of derivations of these results are given in the
Appendix),
\ba
\Delta d_e (\mbox{heavy-heavy}) = e\left({G_F\over
16\pi^2}\right)^2 m_e {\Delta M\over M} {m_1^2\over M}{m_2^2\over
M} M^{-4\ep}
 \left({16\over 3\ep}  - {364\over 9} + {112\over 27}\pi^2 \right) \sin 2\eta.
\label{hh}
\ea
Next, we obtain the contribution when one of $f_i$ is the
light state $\nu$,
\ba \Delta d_e (\mbox{heavy-light}) =
e\left({G_F\over 16\pi^2}\right)^2 m_e {\Delta M\over M}
{m_1^2\over M}{m_2^2\over M} M^{-6\ep}M_W^{2\ep}
 \left(-{16\over 3\ep}  + {104\over 9} \right)\sin 2\eta.
\label{hl}
\ea The sum of both contributions is finite, \ba d_e  &=& \Delta
d_e (\mbox{heavy-heavy}) +\Delta d_e (\mbox{heavy-light})
\nonumber \\
&=& e\left({G_F\over 16\pi^2}\right)^2 m_e {\Delta M\over M}
{m_1^2\over M}{m_2^2\over M}
 \left( {32\over 3}\ln{M\over M_W} - {260\over 9} + {112\over 27}\pi^2 \right)
\sin 2\eta. \label{final}
\ea
Our answer shows that the electron
EDM first appears at $O(m_em_1^2m_2^2M^{-2}G_F^2)$ level, which is
essentially the same as $O(m_em_\nu^2G_F^2)$  if the see-saw relations
$m_{1,2}^2/M\sim O(m_\nu)$ are in place.

In a slightly different model, with two generations of active
neutrinos with masses $m_{\nu1}$ and $m_{\nu2}$ there is an
additional possibility of $f_{1}$ and $f_2$ both being light. It
is easy to see, however, that in the case of (light-light) diagram
the contribution to $d_e$ is parametrically much smaller than Eq.
(\ref{final}), being suppressed by four powers of light neutrino
mass $m_{\nu1}m_{\nu2} (m_{\nu1}^2-m_{\nu2}^2)$. Thus, the
generalization to the case of three generations would not bring
any parametric change to the size of $d_e$. Fixing $m_\nu$, we can
also allow $m_{1,2}$ and $M_{1,2}$ to vary. If $M_{1,2}$ becomes
smaller than the electroweak scale, $d_e$ acquires additional
suppression by $M^2/M_W^2$. This proves that (\ref{final}) with
the chosen hierarchy of scales represents the largest possible
contribution to $d_e$ at fixed $m_\nu$.

\section{Discussion and Conclusions}

What is the largest numerical value of the answer (\ref{final})?
For the see-saw type relation, in the absence of substantial
cancellations between the two terms in (\ref{mnu}), we can
use the following inequality
\ba \left| {m_1^2\over M}{m_2^2\over M}\sin
2\eta\right| \la m_\nu^2.
\label{seesawbound}
\ea
This leads to an extremely tight theoretical bound on the possible size of the EDM,
\ba
|d_e| \la  e\left({G_F\over 16\pi^2}\right)^2 m_e m_\nu^2{|\Delta
M|\over M} \left( 10.7\times\ln{M\over M_W} + 12.1\right),~~~ {\rm see\mbox{-}saw~case}
.
\label{upper}
\ea
Using $|\Delta M| \sim M \sim 10^{16}$ GeV, and
the Particle Data Group book \cite{PDG} bound on the mass of the
electron neutrino, $m_\nu<3$ eV, we arrive at our final numerical
result for the see-saw case,
\ba
|d_e|< 1.5\times 10^{-43} e~{\rm cm}, ~~~ {\rm see\mbox{-}saw~case}.
\label{fnum}
\ea
Not only is this number much smaller than the most optimistic
accuracy of future electron EDM searches \cite{future}, but also
it is significantly smaller than EDMs induced by the KM phase from
the quark sector.

The fine-tuned case deserves special consideration, as the numerical
answer for $d_e$ can be significantly larger than (\ref{fnum}).
Introducing another angle $\delta=\pi/2-\eta$, which has to be small for the
fine-tuning of $m_\nu$ to happen, we rewrite the light neutrino
mass squared as
\ba
m_\nu^2 = \fr{1}{M^2}\left((m_1^2-m_2^2)^2+4m_1^2m_2^2\delta^2\right).
\label{newmnu}
\ea
On the other hand, the answer for EDM contains a factor
${m_1^2m_2^2 \over M^2}\sin(2\eta)=
2{m_1^2m_2^2\over M^2}\delta$, which in view of Eq. (\ref{newmnu})
can be limited as
\ba
\left|\fr{m_1^2m_2^2}{M^2}2\delta\right| < \fr{m_1m_2}{M}m_\nu
\la (0.1 \times 500 ~{\rm GeV})m_\nu.
\label{anotherbound}
\ea
In this expression, we have used (\ref{univ}) and the upper bound
on $m_{1,2}$ taken to be $\sim 500$ GeV. The latter follows from the
condition that Yukawa sector remains perturbative, {\em i.e.}
$y_{1,2}\la O(1-10)$. Comparing (\ref{seesawbound}) and (\ref{anotherbound}),
we observe that the fine-tuned case may lead
to up to $10^{10}$ enhancement relative to the see-saw
case and the electron EDM may reach a value of
\ba
d_e \la 10^{-33}~e~{\rm cm},~~~ {\rm fine\mbox{-}tuned ~ case}.
\label{maximal}
\ea
This value is still far from the existing or projected experimental
accuracy but is much larger than the electron EDM induced by the KM phase.

In some models the effects of the CP-odd electron-nucleon
interaction, $C_S \bar NN \bar e i \gamma_5e$ may dominate over
the electron EDM contribution \cite{Cs} in the atomic (molecular)
EDM. An analysis of possible two-loop diagrams for $C_S$ shows
that in the model with CP violation coming from Majorana neutrinos
this is not the case, and the contribution of $C_S(\eta)$ to
atomic EDM is smaller than (\ref{fnum}) for the see-saw case and
(\ref{maximal}) for the fine-tuned case.

To summarize, we have calculated the contributions of the two-loop
diagrams to the EDM of the electron in non-supersymmetric models
with CP violation in the lepton sector and with Majorana masses
for neutrinos. We notice that the non-zero result for $d_e$ can be
obtained in a rather minimalistic way: with one generation of SM
leptons and two right-handed neutrinos. In terms of the
right-handed mass $M$ and Dirac mass $m_D$, the non-zero result
appears in order $O(m_em_D^4M^{-2}G_F^2)$, which is the largest
possible result given the symmetries of the model. If the
smallness of the light neutrino mass is achieved via usual see-saw
relation without any fine-tuning, this answer is equivalent to
$O(m_em_\nu^2G_F^2)$ and numerically is extremely small. Numerical
smallness of this result stems from the smallness of $m_\nu$, and
the parametric suppression of EDM is very similar to the
suppression of {\em e.g.} the SM amplitude for $\mu\to e\gamma$
decay. In the fine-tuned case, when the smallness of the light
neutrino mass is achieved via near perfect cancellation of two
contributions, the result for $d_e$ may become larger by many
orders of magnitude, but still smaller than about $10^{-33}e~$cm,
and therefore much smaller than the sensitivity of any EDM
experiment in the foreseeable future.

We conclude that the KM-type models with Majorana neutrino masses
do not have any impact on the EDM searches. Therefore, the only
options of searching for CP-violation in the neutrino sector
suggested by the leptogenesis scenario are the CP asymmetries in
neutrino oscillations and not EDMs. Conversely, possible positive
results from the future electron EDM searches could be an
indication of leptogenesis combined with the soft supersymmetry
breaking.

\section*{Acknowledgement}
This research was supported by the Science and Engineering
Research Canada, and by the Collaborative Linkage Grant
PST.CLG.977761 from the NATO Science Programme.

\appendix
\section{Details of the calculation of two-loop diagrams}
Here we give intermediate results for the sum of the  three
diagrams in Fig.~\ref{edm}, before we add and simplify
contributions from $n_{1,2}$.

For the heavy-heavy case we get \ba \Delta d_e
(\mbox{heavy-heavy}) &=& -{ie^5 \over 2m_e s_W^4} {1\over
(16\pi^2)^2 } \left( { m_1 m_2 \over M_1 M_2}\right)^2 \left(
e^{2i\eta} - \mbox{ complex conjugate}\right) \nonumber \\ &&
\cdot \left( M_2 - M_1 \right) {M m_e^2\over M_W^4}M^{-4\ep}
\left( {1\over  6\ep} - {91\over 72}  + {7\over 54}\pi^2 \right)
\nonumber \\
&\simeq & \left({G_F \over 16\pi^2}\right)^2  m_e e{\Delta M m_1^2
m_2^2  \over M^3} M^{-4\ep} \left( {16\over  3\ep} - {364\over 9}
+ {112\over 27}\pi^2 \right)  \sin 2\eta.
\nonumber \\
\label{aphh} \ea

For the heavy-light case we find \ba \Delta d_e
(\mbox{heavy-light}) &=& -{ie^5 \over 2m_e s_W^4} {1\over
(16\pi^2)^2 } {m_e^2 \over M_W^4}
\nonumber \\
&& \hspace*{-30mm}  \cdot \left[ {m_1^2 \over M_1^2}\,e^{2i\eta}
f\left( -{m_2^2 \over M_2}, M_1\right) +{m_2^2 \over M_2^2} \,
f\left( -{m_1^2 \over M_1}e^{-2i\eta}, M_2\right) - \mbox{ complex
conjugate} \right],
\nonumber \\
\ea where the function $f$ arises from the evaluation of the
diagrams without coupling constants, \ba f(m_\nu, M_i) \equiv
m_\nu M_i M_W^{-4\ep}  \left(
 - {1\over 12\ep}\ln {M_i^2\over {M_W^2}}-{5\over 16\ep}
 + {13\over 72}\ln {M_i^2\over {M_W^2}} + {1\over 8}\ln^2 {M_i^2\over {M_W^2}}
  - {31\over 48} + {\pi^2\over 36}
 \right).
\nonumber \\
\ea Here $m_\nu$ and $M_i$ denote a light and a heavy neutrino
mass insertion. Working to first order in the heavy neutrino mass
splitting $\Delta M$, we get \ba \Delta d_e (\mbox{heavy-light})
&=& -{ie^5 \over 2m_e s_W^4} {1\over (16\pi^2)^2 } {m_e^2 \over
M_W^4}
\nonumber \\
&& \cdot {m_1^2 m_2^2 \over M^3} i {\Delta M \over 16}
 M_W^{-4\epsilon}
\left( -{16\over 3\epsilon} +{104\over 9}  + 16 \ln {M^2 \over
M_W^2} \right) \sin 2 \eta
\nonumber \\
&=& \left({G_F\over 16\pi^2}\right)^2
 {m_e e}
{{\Delta M } m_1^2 m_2^2 \over M^3}
 M_W^{-4\epsilon}
\left( -{16\over 3\epsilon} +{104\over 9}  + 16 \ln {M^2 \over
M_W^2} \right) \sin 2 \eta
\nonumber \\
\label{aphl} \ea We see that in the sum of contributions in
\eq{aphh} and \eq{aphl}, the divergences cancel and we obtain the
final result, eq.~\eq{final}.

\section{Asymptotic Expansions}
In this appendix, we show how the  method of asymptotic expansions
\cite{AE} is used to calculate the contributing diagrams for the
electric dipole moment.

Two-loop Feynman diagrams containing more than one mass scale are
difficult to solve analytically.  Analytic expansions in small
parameters, such as the ratio of lepton and weak boson masses, are
used to reduce the calculations to integrals involving only one
mass scale.  The results are integrals which have analytic
solutions.  As an example of this technique, we use Feynman
diagrams from Fig. \ref{edm} containing one light neutrino,
$f_{1}$, of mass $m_{\nu}$ and one heavy neutrino, $f_{2}$, of
mass $M$.

\begin{figure}[htb]
\hspace*{25mm}
\begin{center}
\epsfxsize=45mm \epsfbox{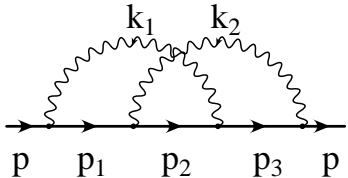} \caption{Momentum assignment
for the EDM calculation.} \label{fig:momentum}
\end{center}\end{figure}

The momentum assignments for the calculation are illustrated in
Fig. \ref{fig:momentum}.  Conservation of momentum at each vertex
requires $p_{1}=p+k_{1}$, $p_{2}=p+k_{1}+k_{2}$ and
$p_{3}=p+k_{2}$.

In order to perform the expansions, all mass scales must be
identified.  For our calculation, the mass scales involved are $M
\gg M_{W} \gg m_{e} \gg m_{\nu}$.

Once the mass scales are determined, the integral volumes are
divided into regions so that the momentum flow through the
internal lines are on the order of one of the mass scales.  With
one light and one heavy neutrino, there are four scenarios to be
considered:
\begin{enumerate}
\item $ k_{1},k_{2}  \ll  M, M_{W}$ \item $ k_{2}      \sim M,
k_{1} \ll M, M_{W} $ \item $ k_{1}, k_{2} \sim M $ \item $
k_{1},k_{2}  \sim M, k_{1}+k_{2} \sim m_{e}$
\end{enumerate}

Within each region, the appropriate propagators can be expanded in
a Taylor series to reduce the number of mass scales in the
integrals.

\subsection{$k_{1},k_{2} \ll M, M_{W}$}

In this momentum region, the heavy neutrino and the $W$ boson
propagators can be expanded in a Taylor series.  For example, the
heavy neutrino propagator can be written as
\begin{equation}
\frac{1}{k^{2}+M^{2}}
    =
\frac{1}{M^{2}}
    \sum_{n=0}^{\infty}
    \left(
        \frac{k^{2}}{M^{2}}
    \right)^{n}
\end{equation}
since $k \ll M$.

\begin{figure}[htb]
\begin{center}
\epsfxsize=45mm \epsfbox{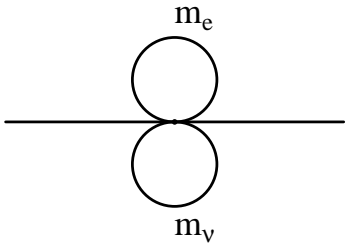} \caption{Reduction of a
two-loop diagram to two one-loop diagrams for $k_{1}, k_{2} \ll M,
M_{W}$.  The solid lines represent massive propagators.}
\label{fig:smallk}
\end{center}
\end{figure}

The Taylor expansion reduces the diagram in Fig.
\ref{fig:momentum} to that of Fig. \ref{fig:smallk}.  The two-loop
diagram is reduced to the product of two one-loop diagrams. Each
one-loop diagram has one mass scale and has an analytic solution.
\subsection{$ k_{2} \sim M, k_{1} \ll M, M_{W} $}

\begin{figure}[htb]
\begin{center}
\epsfxsize=45mm \epsfbox{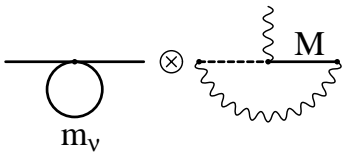} \caption{Reduction
of a two-loop diagram to two one-loop diagrams for $ k_{2} \sim M,
k_{1} \ll M,M_{W} $.  The dashed line represents a massless
propagator.} \label{fig:1large_1small}
\end{center}
\end{figure}

When one momentum is large and the other is small, the propagators
are expanded in all momenta except for the large momentum.  This
reduces the calculation to the diagrams in Fig.
\ref{fig:1large_1small}, a product of two one-loop diagrams.
Clearly, the first diagram has only one mass scale involved.  In
the second diagram, partial fractions are used to further separate
the mass scales.  In particular, we use the identity
\begin{equation}
\frac{1}{k^{2}+M^{2}} \frac{1}{k^{2}+M_{W}^{2}}
    =
\frac{1}{M_{W}^{2}-M^{2}} \left(
    \frac{1}{k^{2}+M^{2}}
    -
    \frac{1}{k^{2}+M_{W}^{2}}
\right)
\end{equation}
The resulting diagrams can be computed analytically.

\subsection{$ k_{1}, k_{2} \sim M $}

\begin{figure}[htb]
\begin{center}
\epsfxsize=45mm \epsfbox{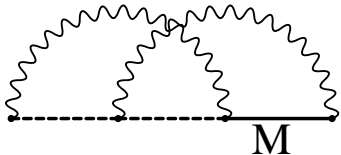} \caption{Taylor Expansion for
$k_{1}, k_{2} \sim M$.} \label{fig:large_k}
\end{center}
\end{figure}

If both momenta are large, a reduction is made to the form of Fig.
\ref{fig:large_k}.  Notice however, that there are still two mass
scales in the diagram.  Thus, we use the assumption that $M \gg
M_{W}$ for further expansion to reduce the diagram down to a one
mass scale diagram.

\subsection{$ k_{1},k_{2}  \sim M, k_{1}+k_{2} \sim m_{e}$}

\begin{figure}[htb]
\begin{center}
\epsfxsize=45mm \epsfbox{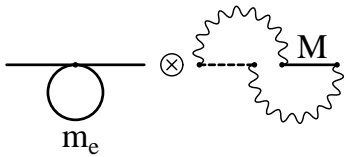} \caption{Taylor expansion
for the case $k_{1}+k_{2} \ll M$.} \label{fig:sum}
\end{center}
\end{figure}

The last case reduces to Fig. \ref{fig:sum}
but does not contribute to the $d_{e}$ in the leading order.

After integration over the initial ranges is performed, the
contributions from all momentum regions are summed to produce eq.
(\ref{hl}).

Using a similar analysis, eq. (\ref{hh}) was computed for the
diagrams with two heavy neutrinos.


\end{document}